\documentclass[a4paper,11pt]{article}
\usepackage{pos}
\usepackage{verbatim}
\graphicspath{ {figures/} }

\newcommand{\Sp}[1]{\ensuremath{\text{Sp}(#1)}}
\newcommand{\SU}[1]{\ensuremath{\text{SU}(#1)}}

\title{Singlet Mesons in Dark Sp(4) Theories}

\author*[a]{Fabian Zierler}
\author[b,c]{Jong-Wan Lee}
\author[a]{Axel Maas}
\author[a]{Felix Pressler}

\affiliation[a]{Institute of Physics, NAWI Graz, University of Graz, Universitätsplatz 5, 8010 Graz, Austria}
\affiliation[b]{Department of Physics, Pusan National University, Busan 46241, Korea}
\affiliation[c]{Institute for Extreme Physics, Pusan National University, Busan 46241, Korea}

\emailAdd{jwlee823@pusan.ac.kr}
\emailAdd{axel.maas@uni-graz.at}
\emailAdd{f.pressler@edu.uni-graz.at}
\emailAdd{fabian.zierler@uni-graz.at}

\abstract{We explore some aspects of $Sp(4)$ gauge theory with two fundamental fermions in the context of composite Goldstone Dark Matter. We present preliminary lattice results for the mass of the pseudoscalar iso-singlet meson $\eta'$ using unimproved Wilson fermions and the standard plaquette action. We find that the $\eta'$ is slightly heavier than the pseudoscalar non-singlets $\pi$ and lighter than the vector mesons $\rho$ for multiple ensembles with $m_\pi/m_\rho \geq 0.77$. This pattern is potentially relevant for \textit{Beyond the Standard Model} physics model building. Furthermore, we show that for $N_f=1+1$ flavours the disconnected contributions to the unflavoured pseudoscalar $\pi^0$ are small. We supplement this measurement by a calculation of the $\pi\pi$ scattering length $a_0$ which shows that the ensemble studied might be phenomenologically relevant for Dark Matter models such as the \textit{Strongly Interacting Massive Particles} paradigm.}

\FullConference{%
The 39th International Symposium on Lattice Field Theory,\\
8th-13th August, 2022,\\
Rheinische Friedrich-Wilhelms-Universität Bonn, Bonn, Germany
}
\begin{document}
\maketitle

\section{Introduction}
The structure of symplectic gauge groups $\Sp{2N}$ with fermions has provided interesting opportunities for physics beyond the Standard Model (BSM). For a sufficiently small number of fermions they are asymptotically free, confining and exhibit (spontaneous) chiral symmetry breaking as in QCD and unitary gauge groups in general. Since the fundamental representation of Dirac fermions is pseudo-real, however, their global symmetries differ from those of a unitary gauge group $\SU{N_c\geq 3}$. For $N_f$ flavors it leads to the symmetry breaking pattern $\SU{2N_f} \rightarrow \Sp{2N_f}$ \cite{Kosower:1984aw}, which is of considerable interest for BSM physics. In the case of $N_f=2$, this provides a realization of the $\SU{4}/\Sp{4}$ coset for composite Higgs models (see e.g. Ref.~\cite{Cacciapaglia:2020kgq} for a review). Furthermore, the resulting five (pseudo-)Goldstone bosons play the role of {\it Strongly Interacting Massive Particles} (SIMP) in the context of dark matter (DM), where a large number of colours appear to be phenomenologically more relevant \cite{Hochberg:2014kqa} motivating lattice investigations beyond $\SU{2}$. 

The spectrum of flavour-singlet mesons is particularly interesting for DM models as they are not protected by flavour symmetry from decays into Standard Model (SM) particles. Of particular interest is the pseudoscalar singlet $\eta'$ which is associated with the $U_A(1)$ axial anomaly. In $\SU{3}$ gauge theory with two flavours this state is lighter than the vector mesons \cite{Dimopoulos:2018xkm} and is expected to be degenerate with the other non-singlet pseudoscalar mesons in the $N_c \to \infty$ limit. 

On the lattice the non-singlet spectrum of mesons in the $\Sp{4}$ two-flavour theory has been studied in quenched approximation \cite{Bennett:2019cxd}, with dynamical degenerate fermions \cite{Bennett:2019jzz} and with dynamical non-degenerate fermions \cite{Kulkarni:2022bvh} (see also Ref.~\cite{Bennett:2021mbw} for a recent review on $\Sp{2N}$ lattice gauge theory). Similar studies have been performed in $\SU{2}$ gauge theory ( e.g. Ref.~\cite{Arthur:2016dir}) and first investigations on the spectrum of singlet mesons have been carried out \cite{Arthur:2016ozw}. In this contribution we present first results on the $\eta'$ meson in $\Sp{4}$ gauge theory with two fundamental flavours for degenerate and non-degenerate fermions. We also report on the disconnected contributions to the the unflavoured pseudoscalar $\pi^0$ meson in the non-degenerate case. In a SIMP scenario the fermions need to be massive to provide a pseudo-Goldstone DM model. However, it is unclear how heavy the fermions have to be in order for the theory to be a viable DM candidate. We therefore perform an exploratory investigation of the length of $\pi\pi$ scattering in the isospin $I=2$ channel along the lines of a similar study in $\SU{2}$ gauge theory \cite{Arthur:2014zda} and match the results to experimental constraints. 

\section{Lattice setup}
We study four-dimensional \Sp{4} gauge theory with two fundamental fermions on the Euclidean lattice using the Wilson plaquette action and unimproved Wilson fermions. All our calculations have been performed with the HiRep code \cite{DelDebbio:2008zf} extended to support symplectic gauge theories \cite{Bennett:2017kga, githubGitHubSa2cHiRep}. To generate gauge configurations we use the Hybrid Monte Carlo (HMC) algorithm for degenerate fermions and the rational HMC (RHMC) algorithm for non-degenerate ones. In the latter case we monitored the lowest eigenvalue of the Dirac operator as its determinant is not guaranteed to be positive definite. For the masses considered in this work we do not observe any hints of a sign problem \cite{Kulkarni:2022bvh}, where the details of the ensembles are found in Table \ref{tab:masses}.

\subsection{Pseudoscalar Mesons}
For $N_f=2$ with degenerate fermions mesons appear either in $10$-plets, $5$-plets or as singlets under the global $\Sp{4}_F$ flavour symmetry which arises after the symmetry breaking through fermion condensates and/or fermion masses \cite{Kosower:1984aw,Drach:2017btk}. The theory has a $5$-plet of pseudoscalar mesons which are the (pseudo-)Goldstones of global symmetry breaking, and an additional flavour singlet $\eta'$ which is connected to the axial anomaly. When introducing strong isospin breaking by setting $m_u \neq m_d$, the $5$-plet decomposes into a $4$-plet of flavoured mesons and a flavour singlet $\pi^0$ under the remaining $\SU{2}_u \times \SU{2}_d$ global flavour symmetry. Following the standard procedure we extract their masses from the meson propagator $\langle O (n) \bar O(m) \rangle$. Here, $n,m$ are labels for lattice sites. We are interested in operators of the form 
\begin{align}
    O_1(n) &= \bar u \Gamma d, \\
    O_2(n) &= \left( \bar u \Gamma u - \bar d \Gamma d \right) / \sqrt{2}, \\
    O_3(n) &= \left( \bar u \Gamma u + \bar d \Gamma d \right) / \sqrt{2}, 
\end{align}
where $u$ and $d$ denote two distinct Dirac flavours and $\Gamma$ is an appropriate gamma structure. For $\Gamma=\gamma_5$ these operators probe the pseudoscalar mesons. The four flavoured pseudoscalars $\pi^\pm$ are probed by $O_1$, $\pi^0$ by $O_2$, and the flavour singlet $\eta'$ by $O_3$. The difference between the $\pi^0$ and $\eta'$ arises in the form of disconnected diagrams. Physically, they correspond to pure gluonic propagation. In the isospin-symmetric limit these contributions cancel for $\pi^0$ and its non-singlet nature for $m_u = m_d$ becomes manifest. 

Other singlet states exist for any $\Gamma$ where the non-singlets for degenerate fermions appear in a $5$-plet. However, no singlet states appear for other gamma structures, such as $\Gamma=\gamma^\mu$ interpolating the vector mesons $\rho$. States involving more fermions and/or additional gluons are still possible. The state that would source the $\phi/\omega$ meson in QCD is part of the $\rho$ $10$-plet and under isospin breaking the $10$-plet decomposes into a flavoured $4$-plet and an unflavoured $6$-plet under the remaining $\SU{2}_u \times \SU{2}_d$ symmetry. Another singlet of interest is the scalar flavour-singlet which corresponds to the $\sigma /f_0$ in QCD. This meson has the same quantum numbers as the vacuum. Therefore, its correlator contains an additional constant several orders of magnitude larger than the exponentially decaying signal with the ground state energy, the mass of the scalar meson, at late times. As an adequate subtraction of this constant is needed, we defer the study of the scalar flavour-singlet to future work. 

In general, the disconnected contributions suffer more strongly from noise and thus higher statistics is needed. Additionally, they require all-to-all propagators. We use noisy \cite{Bitar:1988bb}, diluted \cite{Foley:2005ac} sources and find that spin dilution and even-odd dilution on $Z_2 \times Z_2$ noise sources with $n_\text{hit}=128$ are sufficient to obtain a signal for the pseudoscalar disconnected diagrams for the ensembles listed in Table \ref{tab:masses}. From this we build an unbiased estimator for the disconnected pieces as outlined in Ref.~\cite{Arthur:2016ozw}.

\begin{figure}
    \centering
    \includegraphics[width=0.48\textwidth]{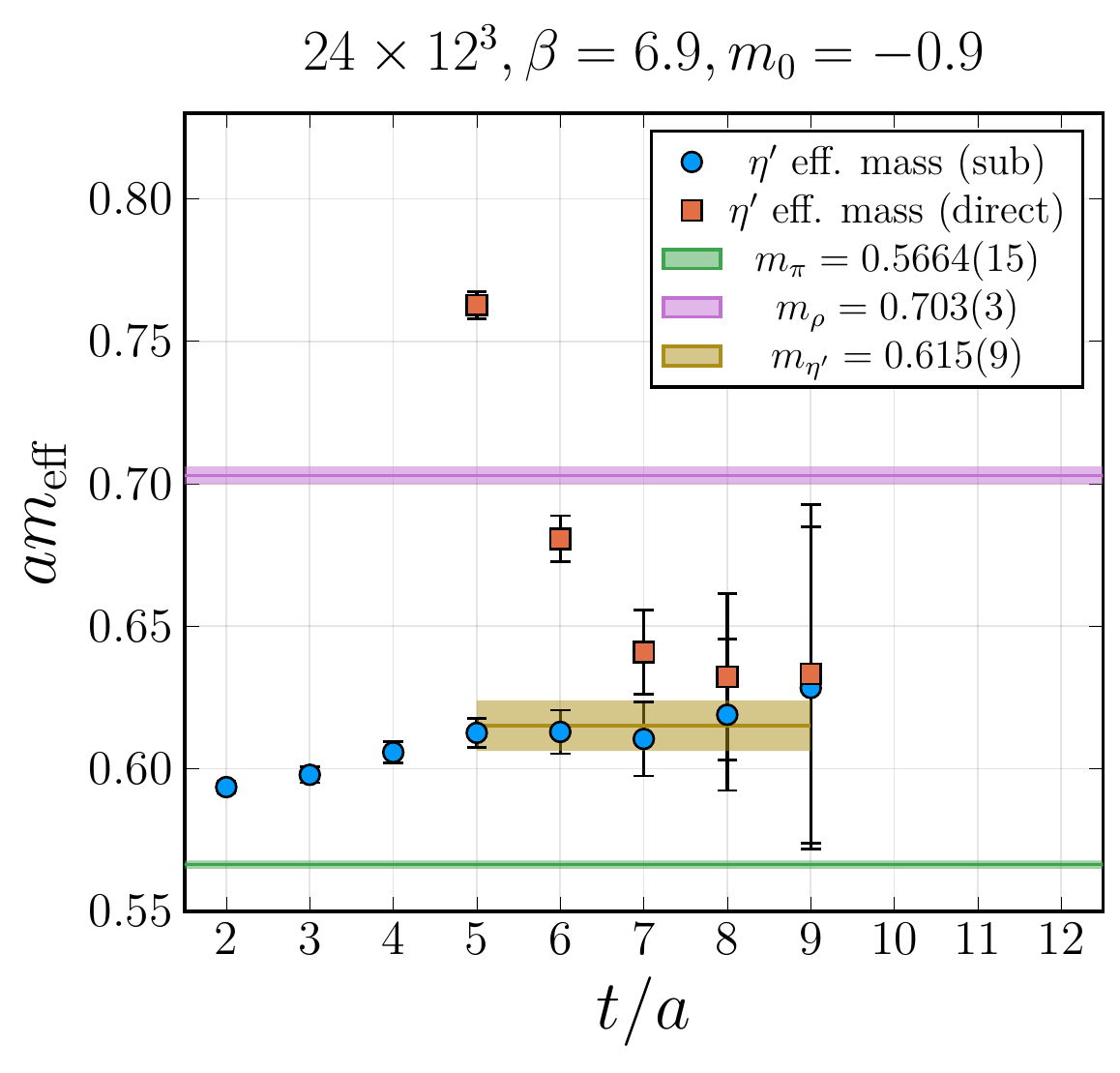}
    \includegraphics[width=0.49\textwidth]{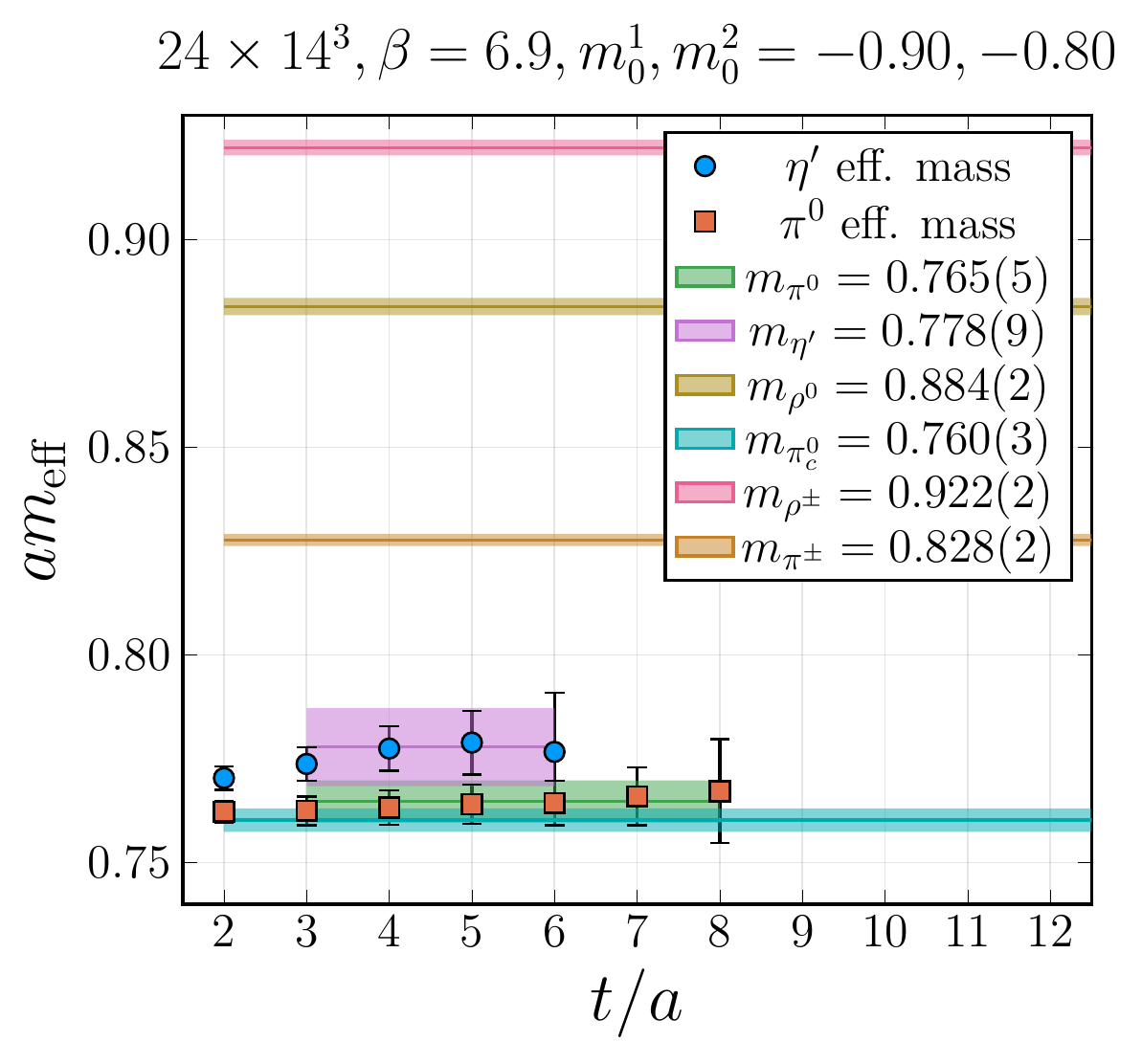}
    \caption{(left) Effective masses of the $\eta'$ correlator before subtraction of excited state contributions (orange squares) and after subtraction (blue circles). We only show data points for which we can extract a signal. The subtracted effective mass shows a clear plateau over several timeslices. (right) Effective masses for the pseudoscalar singlets $\pi^0$ and $\eta'$ in the $N_f=1+1$ theory after excited state subtraction. Additionally, we give the masses for the unflavoured vector mesons and the $\pi^0$ in the connected-only approximation denoted by $m(\pi^0_c)$.}
    \label{fig:eff_mass}
\end{figure}

The disconnected contributions still suffer from substantial noise and we only find a signal for a few timeslices at small and intermediate $t$. However, it has been noted that the they receive relatively small corrections from the excited state contamination compared to the connected diagrams. Hence, we first remove the excited state contributions present in the connected pieces. This is done by fitting the connected pieces -- which have a good signal over all timeslices -- at large $t$ by a single exponential of the form
\begin{align}
    C_{\pi,\text{conn}}^\text{1exp}(t) &= A_0 \left( e^{-m_\text{conn}t} + e^{-m_\text{conn}(T-t)}  \right),
\end{align}
and use this to build an improved correlator for the $\eta'$ and the $\pi^0$ \cite{Neff:2001zr}
\begin{align}
 	C_{\eta'}^\text{1exp}(t) &= C_{\pi,\text{conn}}^\text{1exp}(t) + C_{\eta',\text{disc.}}(t), \\
 	C_{\pi^0}^\text{1exp}(t) &= C_{\pi,\text{conn}}^\text{1exp}(t) + C_{\pi^0,\text{disc.}}(t).
\end{align}
With the excited, connected contributions removed we obtain plateaus in the effective masses at much earlier time $t$. In the left panel of Fig.~\ref{fig:eff_mass} we depict the effective masses of the singlet mesons with and without excited state subtraction in the connected pieces for an exemplified case of degenerate fermions. From the plateaus in the effective mass we determine the range of $t$ for which we perform a single-exponential fit to the improved correlators. The masses extracted using this method are found in Table~\ref{tab:masses}.
Similarly, we apply the subtraction method to the non-degenerate case for both pseudoscalar singlets $\pi^0$ and $\eta^\prime$ as shown in the right panel of Fig.~\ref{fig:eff_mass}. 
\begin{table}
    \setlength{\tabcolsep}{3pt}
    \begin{tabular}[t]{|c|c|c|c|c|c|c|c|c|c|c|c|}
        \hline \hline 
        $\beta$ & $m_0$ & $L$ & $T$ & $n_\text{conf}$ & $m_\pi$ & $m_\rho$    & $m_{\eta'}$ \\
        \hline \hline 
            6.9 & -0.87 & 12  & 24  & 309  & 0.743(3)  & 0.852(4)   & 0.780(13)   \\
            6.9 & -0.89 & 12  & 24  & 575  & 0.631(3)  & 0.753(5)   & 0.669(14)   \\
            6.9 & -0.90 & 12  & 24  & 2767 & 0.5664(15)& 0.703(3)   & 0.615(9)    \\
            6.9 & -0.90 & 14  & 24  & 932  & 0.5642(19)& 0.698(4)   & 0.614(12)   \\
            6.9 & -0.91 & 14  & 24  & 334  & 0.484(6)  & 0.624(12)  & 0.57(3)     \\
        \hline \hline 
    \end{tabular}
    \\
    \begin{tabular}[t]{|c|c|c|c|c|c|c|c|c|c|c|c|}
        \hline \hline 
        $\beta$ & $m_0^1$ & $m_0^2$ & $L$ & $T$ & $n_\text{conf}$ & $m_{\pi^0_c}$ & $m_{\pi^0}$ & $m_{\pi^\pm}$ & $m_{\rho^0}$ & $m_{\rho^\pm}$ & $m_{\eta'}$ \\ 
        \hline \hline
        6.9 & -0.90 & -0.75 & 14 & 24 & 599 & 0.804(2) & 0.812(4) & 0.918(2) & 0.928(2) & 1.003(2) & 0.821(7) \\ 
        6.9 & -0.90 & -0.80 & 14 & 24 & 599 & 0.760(3) & 0.765(5) & 0.828(2) & 0.884(2) & 0.922(2) & 0.778(9) \\ 
        6.9 & -0.90 & -0.85 & 14 & 24 & 291 & 0.691(2) & 0.691(3) & 0.714(2) & 0.811(2) & 0.822(3) & 0.719(6) \\ 
        6.9 & -0.90 & -0.89 & 14 & 24 & 255 & 0.597(3) & 0.600(5) & 0.600(4) & 0.724(5) & 0.728(7) & 0.63(1)  \\ 
        \hline \hline 
    \end{tabular}
    \caption{Lattice ensembles and meson masses in lattice units for the $N_f=2$ theory (upper) and for strong isospin breaking $N_f=1+1$ (lower). We denote the bare fermion masses by $m_0^{(1,2)}$ and the inverse gauge coupling by $\beta$. All simulations were performed on lattices of the size $L^3 \times T$. For reference we also give the mass of the unflavoured pseudoscalar in the connected-only approximation in the $N_f=1+1$ theory denoted by $m_{\pi^0_c}$. }
    \label{tab:masses}
\end{table}
\begin{figure}
    \centering
    \includegraphics[width=0.48\textwidth]{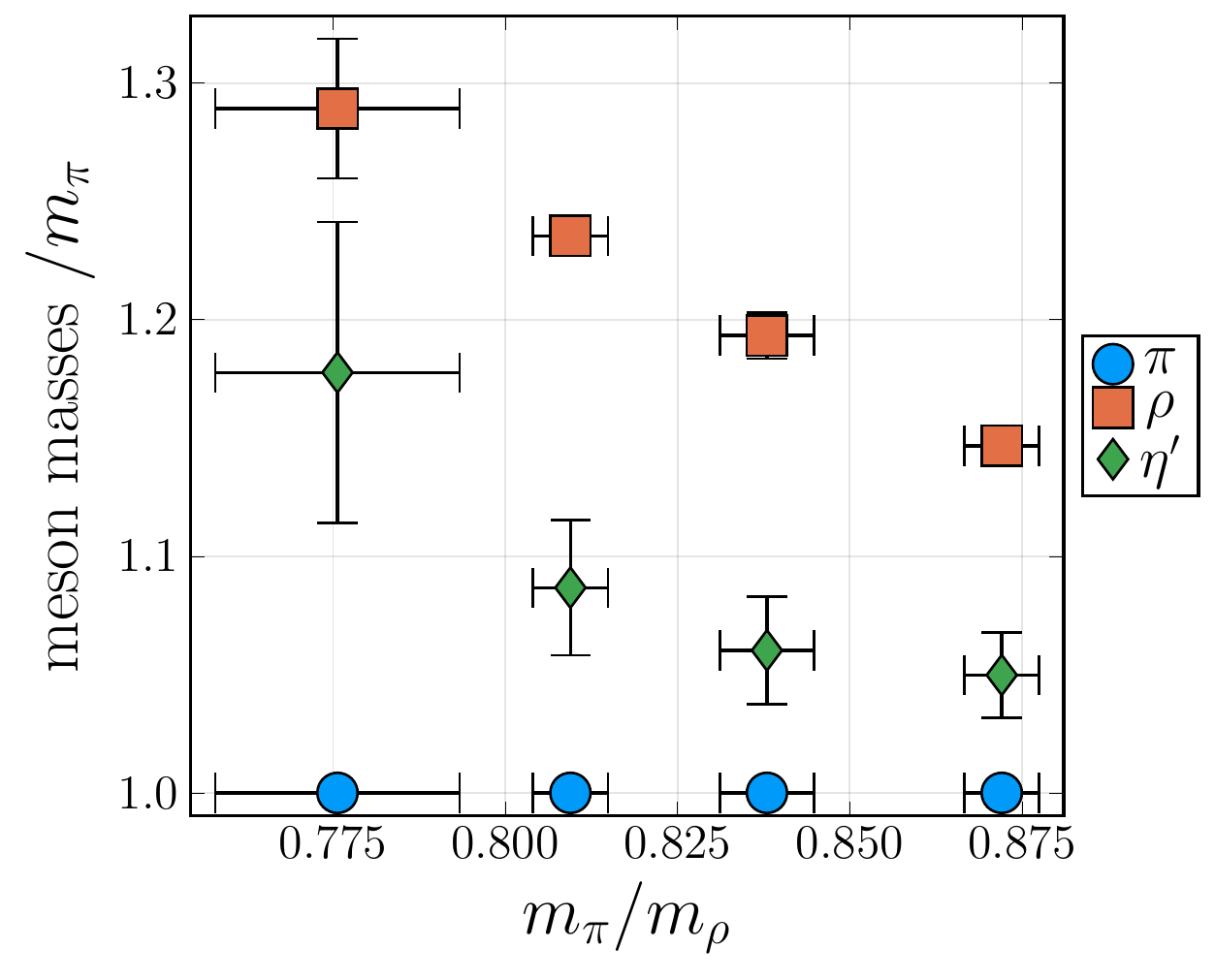}
    \includegraphics[width=0.50\textwidth]{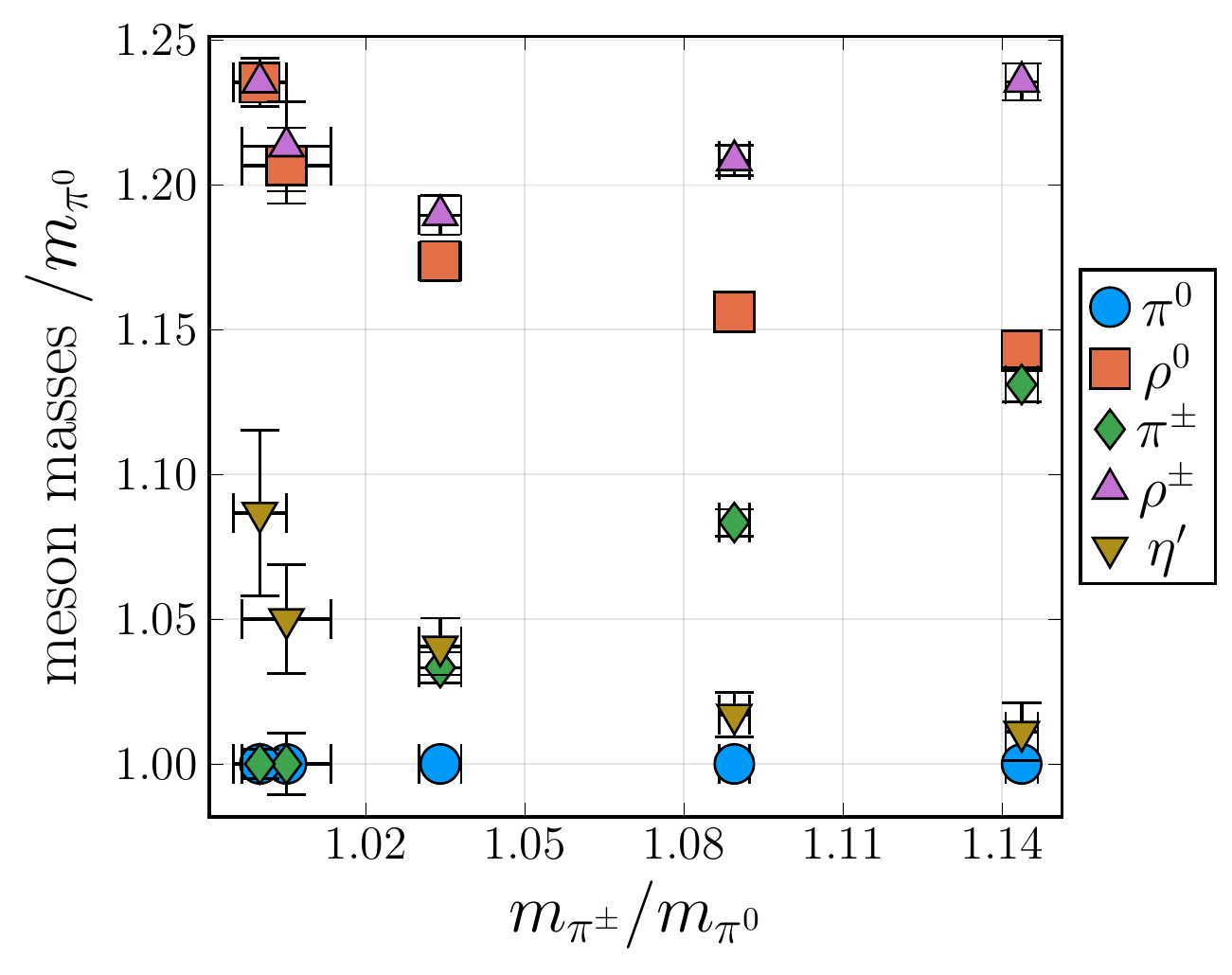}
    \caption{Masses of the pseudoscalar and vector mesons in $N_f=2$ (left) and in $N_f=1+1$ (right) in units of the lightest pseudoscalar meson mass. Note, that in the latter case $\pi^\pm$ and $\rho^\pm$ denote the $4$-plet of flavoured mesons. Similarly, $\rho^0$ denotes the $6$-plet of unflavoured vector mesons.}
    \label{fig:mass}
\end{figure}

We plot the spectrum of pseudoscalar and vector mesons for both the degenerate and non-degenerate case in Fig.~\ref{fig:mass}. We see that for the large fermion masses the mass of $\eta^\prime$ is slightly larger than that of $\pi$. Furthermore, we observe that for non-degenerate fermions the disconnected contributions to the $\pi^0$ meson is small and the connected-only approximation of Ref.~\cite{Kulkarni:2022bvh} appears to be justified in this mass-regime. 
However, it is a priori unknown if the parameter space studied here is able to reproduce the Dark Matter self-scattering rates and Dark Matter masses needed in a SIMP scenario \cite{Kulkarni:2022bvh}. In the next section, we address this by performing an exploratory calculation of the $\pi\pi$ scattering lengths and match it to existing experimental constraints. 

\section{Dark Matter self-scattering: \texorpdfstring{$\pi \pi$}{pi pi} scattering lengths at isospin \texorpdfstring{$I=2$}{I=2}}

Dark Matter self-scattering might address some issues such as the core-cusp problem and strongly interacting Dark Matter models naturally provide self-interactions. Together with the current experimental constraints on the self-interaction of DM we perform a preliminary test on the viability of our parameter choices by computing an estimate of the $\pi\pi$ scattering length $a_0$ in the isospin $I=2$ channel.

A similar, more in-depth study has already been performed in $\text{SU}(2)$ \cite{Arthur:2014zda}. We follow their approach. Since we are (currently) only interested in a order of magnitude estimation, we defer a full Lüscher analysis for future work. We study the ensemble with the highest available statistics given by the bare gauge coupling $\beta=6.9$ and a degenerate, bare fermion mass $m_0 = -0.9$ on a $24 \times 12^3$ lattice. We extract the energy shift $\delta E_{\pi\pi}$ from \cite{Arthur:2014zda}
\begin{align}
   R(t) = \frac{C_{\pi\pi}(t) - C_{\pi\pi}(t+1)}{C_{\pi}^2(t) - C_{\pi}^2(t+1)},
\end{align}
by fitting it to its large time behaviour of 
\begin{align}
   R(t \to \infty) = A \left[ \cosh \left( \delta E_{\pi\pi} (t - T/2) \right) +  \sinh \left( \delta E_{\pi\pi} (t - T/2) \right)  \coth \left( m_{\pi} (t - T/2) \right) \right].
\end{align}
Here, $C_\pi$ and $C_{\pi\pi}$ denote the one- and two-pseudoscalar correlation functions, respectively. From this we estimate the scattering length according to the relation
\begin{align}
    \frac{\delta E_{\pi\pi}}{m_\pi} = \frac{4\pi m_\pi a_0}{(m_\pi L)^3} \left( 1 + c_1 \frac{m_\pi a_0}{m_\pi L} +  c_2 \left( \frac{m_\pi a_0}{m_\pi L}\right)^2 \right)
\end{align}
with coefficients $c_1=-2.837$ and $c_2=6.375$ \cite{Luscher:1986pf}. We obtained values of $\delta E_{\pi\pi} = 0.010(4)$ and $a_0=1.0(5)$. Using $\sigma = \pi a_0^2$ we compare our result with experimental constraints on $\sigma / m_\pi^{phys.}$. Note that we have not yet fixed the physical lattice spacing $a$. From the experimental constraints $\sigma / m_\pi^{phys.} < 0.19 ~ \text{cm}^2g^{-1}$ \cite{Eckert:2022qia} and $\sigma / m_\pi^{phys.} < 0.13 ~ \text{cm}^2g^{-1}$ \cite{Andrade:2020lqq} we estimate a lower limit for the physical dark pion mass and obtain $m_\pi > 100 \text{MeV}$. This is consistent with the constraints of the Dark Matter relic density \cite{Hochberg:2014dra} and thus provides a hint that we are looking at phenomenologically interesting points in parameter space.

Still, we note that this is a rough estimate and a full scattering analysis is needed. Furthermore, it has been obtained only on a fixed volume for one lattice spacing and one (degenerate) fermion mass. It was pointed out in Ref.~\cite{Arthur:2014zda} that finite volume effects are expected to be large in this system. We also note that the current constraints \cite{Andrade:2020lqq,Eckert:2022qia} are so strong that the core-vs-cusp problem might not be resolved by a velocity-independent cross-section but that rather an explicit velocity dependence $\sigma(v)$ is needed. This further motivates going beyond a study of the scattering length to fully velocity-dependent $\pi\pi$ scattering. 

\section{Summary}
We provide preliminary results on the spectrum of pseudoscalar and vector mesons, including the flavour-singlet, in Sp($4$) gauge theory with $N_f=2$ and $N_f=1+1$ Dirac fermions. We find that for ensembles with $m_\pi / m_\rho > 0.77$ the pseudoscalar singlet $\eta'$ is parametrically close to the non-singlet pseudoscalars $\pi$. In the non-degenerate case we find that for the large quark masses considered here, the connected-only approximation for the $\pi^0$ appears to be justified. We supplement these spectroscopic investigations by a first estimate of the $\pi\pi$ scattering length. Combined with current experimental constraints on Dark Matter self-interaction we find that the ensembles studied in this work warrant further investigations. 

\acknowledgments
FZ is supported by the Austrian Science Fund research teams grant STRONG-DM (FG1). The work of J.~W.~L is supported by the National Research Foundation of Korea (NRF) grant funded by the Korea government (MSIT) (NRF-2018R1C1B3001379). The computations have been performed on the Vienna Scientific Cluster (VSC4). We thank E.~Bennett, H.~Hsiao, S.~Kulkarni, B.~Lucini, S.~Mee and M.~Piai for helpful discussions.

\bibliographystyle{JHEP}
\bibliography{bib.bib}

\end{document}